\documentclass[aps,prl,twocolumn,superscriptaddress,hyperref,showpacs]{revtex4-1}
\usepackage{graphicx}
\usepackage{amsfonts}
\usepackage{natbib}
\usepackage{hyperref}

\begin{document}


\title{A Correlated Electron-Nuclear Dynamics with Conditional Wave Functions}


\author{Guillermo Albareda}
\email[]{albareda@fhi-berlin.mpg.de}
\author{Heiko Appel}
\affiliation{Fritz-Haber-Institut der Max-Planck-Gesellschaft, Faradayweg 4-6, D-14195 Berlin, Germany}
\author{Ignacio Franco}
\affiliation{Department of Chemistry, University of Rochester, Rochester, New York 14627, USA}
\author{Ali Abedi}
\affiliation{Max-Planck Institut f\"ur Mikrostrukturphysik, Weinberg 2, D-06120 Halle, Germany}
\author{Angel Rubio}
\email[]{angel.rubio@ehu.es}
\affiliation{Fritz-Haber-Institut der Max-Planck-Gesellschaft, Faradayweg 4-6, D-14195 Berlin, Germany}
\affiliation{Nano-Bio Spectroscopy group and ETSF Scientific Development Centre, Universidad del Pa\'is Vasco,
CFM CSIC-UPV/EHU-MPC and DIPC, Av. Tolosa 72, E-20018 Donostia, Spain and European Theoretical Spectroscopy Facility}


\date{\today}

\begin{abstract}
The molecular Schr\"odinger equation is rewritten in terms of non-unitary equations of motion for the nuclei (or electrons) that depend parametrically on the configuration of an ensemble of 
generally defined electronic (or nuclear) trajectories. 
This scheme is exact and does not rely on the tracing-out of degrees of freedom. Hence, the use of trajectory-based statistical techniques can be exploited to circumvent 
the calculation of the computationally demanding Born-Oppenheimer potential-energy surfaces and non-adiabatic coupling elements. 
The concept of potential-energy surface is restored by establishing a formal connection with the exact factorization of the full wave function. 
This connection is used to gain insight from a simplified form of the exact propagation scheme.

\end{abstract}


\maketitle
In order to describe the correlated motion of electrons and nuclei, many strategies have been proposed to transcend the picture where the nuclei evolve on top of a 
single Born-Oppenheimer potential-energy surface (BOPES) \cite{TullyPerspective}. 
Using a time-independent basis-set expansion of the electron-nuclear wave function, full quantum studies provide a complete description of non-adiabatic dynamics \cite{domcke}. 
The scaling of these methods (even for a time-dependent basis-set expansion \cite{MCTDH}) is however limiting their use to describe a few degrees of freedom. 
The so-called direct dynamics techniques attempt to alleviate this problem by calculating the BOPESs on-the-fly \cite{robb,*truhlar}.
Of particular interest here are those methods that use information from quantum chemistry or time-dependent density functional theory calculations in the form of forces. 
Ab-initio surface hopping, Ehrenfest dynamics \cite{tully1998mixed}, or Gaussian wavepacket methods (such as the multiple spawning method) \cite{SawadaMetiu,*BurghardtCederbaum,*Martinez}, 
are all able to reproduce the dynamics of some systems of interest \cite{results1,*results2,*results3}.
In most of these methods, however, the form of the nuclear wave function is restricted as they use a local- or classical trajectory-based representation of the nuclear wavepacket.  
In addition to the difficulties of including external fields or calculating the non-adiabatic coupling elements (NACs), this introduces the problem of systematically accounting for quantum nuclear effects.

In this Letter we propose an exact propagation scheme aimed at the study of non-adiabatic dynamics in the presence of arbitrary external electromagnetic fields.
The coupled electron-nuclear dynamics is separated without tracing-out degrees of freedom, which lends to
a rigorous starting point for systematically including non-adiabatic nuclear effects without relying on the computation of BOPESs and NACs. 
This work constitutes a multi-component extension of the conditional formalism proposed in \cite{PRLOriols,PRBGuille,*JCE,*FNLGuille}. Further, the propagation scheme presented here 
generalizes the conditional formalism beyond its original hydrodynamic formulation \cite{PRLOriols}. This makes it suitable to be coupled with well established electronic structure methods. 

Throughout this Letter we use atomic units, and electronic and nuclear coordinates are collectively denoted by $\mathbf{r} = \{ \mathbf{r}_1,...,\mathbf{r}_{N_e} \}$ and 
$\mathbf{R} = \{ \mathbf{R}_1,...,\mathbf{R}_{N_n} \}$, being $N_e$ and $N_n$ respectively the total number of electrons and nuclei. 
The full (non-relativistic) electron-nuclear wave function $\Psi(\mathbf{r},\mathbf{R},t)$ satisfies the TDSE, 
\begin{equation} \label{Scho}
i \partial_t \Psi(t) = \Big\{ \hat T_e(\mathbf{r}) + \hat T_n(\mathbf{R}) +  \hat W(\mathbf{r},\mathbf{R},t) \Big\} \Psi(t),
\end{equation}
where $\hat T_e=\sum\nolimits_{\xi=1}^{N_e} {(-i{\nabla}_\xi - \textbf{A}^{\mathcal{EM}}(\mathbf{r}_\xi))^2}/{2m}$ and 
$\hat T_n=\sum\nolimits_{\nu=1}^{N_n} {(-i\nabla_\nu - \textbf{A}^{\mathcal{EM}}(\mathbf{R}_\nu))^2}/{2M_\nu}$ 
are the electronic and nuclear kinetic energy operators, and $\textbf{A}^{\mathcal{EM}}$ is the external vector potential in the Coulomb gauge due to an arbitrary external electromagnetic field. 
All scalar potentials are included in $\hat W(\mathbf{r},\mathbf{R},t) =  \hat V_{int}(\mathbf{r},\mathbf{R}) + V^e_{ext}(\mathbf{r},t) + \hat V^n_{ext}(\mathbf{R},t)$, where 
$V^e_{ext}$ ($V^n_{ext}$) is the electronic (nuclear) external scalar potential, and 
$\hat V_{int} = \hat W_{ee}(\mathbf{r}) + \hat W_{nn}(\mathbf{R}) + \hat W_{en}(\mathbf{r},\mathbf{R})$ accounts for the internal Coulombic interactions. 
Next, we present the main result of this Letter.

\textit{Theorem.---}
(a) The molecular wave function $\Psi(\mathbf{r},\mathbf{R},t)$ satisfying the TDSE (\ref{Scho}) can be exactly decomposed either in terms of nuclear or electronic
conditional wave functions
\begin{equation}\label{psin}
\psi_{n}(\mathbf{R},t;\mathbf{r}^\alpha(t)) :=
 \int \delta(\mathbf{r}^\alpha(t) - \mathbf{r}) \Psi(\mathbf{r},\mathbf{R},t) d\mathbf{r},
\end{equation}\vspace{0 mm}
\begin{equation}\label{psie}
\psi_{e}(\mathbf{r},t;\mathbf{R}^\alpha(t)) :=
 \int \delta(\mathbf{R}^\alpha(t) - \mathbf{R}) \Psi(\mathbf{r},\mathbf{R},t) d\mathbf{R},
\end{equation}
provided that the ensemble of trajectories $\{ \mathbf{r}^\alpha(t),\mathbf{R}^\alpha(t) \}$ explores the support of $|\Psi(\mathbf{r},\mathbf{R},t)|^2$ 
at any time $t$. 

(b) The conditional wave functions $\psi_{n}(\mathbf{R},t;\mathbf{r}^\alpha(t))$ and $\psi_{e}(\mathbf{r},t;\mathbf{R}^\alpha(t))$ obey respectively the following non-unitary equations of motion:
\begin{widetext}
\begin{equation}\label{cond_2}
i d_t\psi_{n}(\mathbf{R},t;\mathbf{r}^\alpha(t))  = \Big\{  \hat T_n  +  \hat W(\mathbf{r}^\alpha(t),\mathbf{R},t) \Big\} \psi_{n}(\mathbf{R},t;\mathbf{r}^\alpha(t)) 
\nonumber\\
+  {\hat T_e \Psi(\mathbf{r}, \mathbf{R} ,t)} \big|_{\mathbf{r}^\alpha(t)}    
\nonumber \\
+ i {\mathbf{\nabla}_{\mathbf{r}} \Psi(\mathbf{r}, \mathbf{R} ,t)} \big|_{\mathbf{r}^\alpha(t)} \cdot \dot{ \mathbf{r}}^\alpha(t),
\end{equation}
\begin{equation}\label{cond_1}
i d_t\psi_{e}(\mathbf{r},t;\mathbf{R}^\alpha(t))  = \Big\{  \hat T_e  +  \hat W(\mathbf{r},\mathbf{R}^\alpha(t),t) \Big\} \psi_{e}(\mathbf{r},t;\mathbf{R}^\alpha(t)) 
\nonumber\\
+  {\hat T_n \Psi(\mathbf{r}, \mathbf{R} ,t)} \big|_{\mathbf{R}^\alpha(t)}    
\nonumber \\
+ i {\mathbf{\nabla}_{\mathbf{R}} \Psi(\mathbf{r}, \mathbf{R} ,t)} \big|_{\mathbf{R}^\alpha(t)} \cdot \dot{ \mathbf{R}}^\alpha(t).
\end{equation}
\end{widetext}
For the sake of simplicity, we omit from now on the explicit time-dependence of the trajectories, 
i.e. $\left\{\mathbf{r}^{\alpha},\mathbf{R}^{\alpha}\right\} \equiv \left\{\mathbf{r}^{\alpha}(t),\mathbf{R}^{\alpha}(t)\right\}$.

\textit{Proof.---} Part (a) To demonstrate that expressions (\ref{psin}) and (\ref{psie}) are exact decompositions of the molecular wave function, we only need to realize that 
an ensemble of these conditional wave functions can be used to reconstruct the full wave function as follows,  
\begin{equation}\label{reconstruction_n}
\Psi(\mathbf{r}, \mathbf{R} ,t) = 
  \left\{ \begin{array}{ll}
	    \hat{\mathcal{D}}_\mathbf{r} \left[\psi_{n}\right], & \mbox{if } \sum_{\alpha=1}^{\infty}\delta (\mathbf{r}^\alpha - \mathbf{r}) \neq 0 \\
	    0, & \mbox{if } \sum_{\alpha=1}^{\infty}\delta (\mathbf{r}^\alpha - \mathbf{r}) = 0  
	   \end{array}
  \right.
\end{equation}
or
\begin{equation}\label{reconstruction_e}
\Psi(\mathbf{r}, \mathbf{R} ,t) = 
  \left\{ \begin{array}{ll}
	    \hat{\mathcal{D}}_\mathbf{R} \left[\psi_{e}\right], & \mbox{if } \sum_{\alpha=1}^{\infty}\delta (\mathbf{R}^\alpha - \mathbf{R}) \neq 0 \\
	    0, & \mbox{if } \sum_{\alpha=1}^{\infty}\delta (\mathbf{R}^\alpha - \mathbf{R}) = 0    
          \end{array}
  \right.
\end{equation}
where we have defined the transformations, $\hat{\mathcal{D}}_\mathbf{r} [f(\mathbf{r}^\alpha)] \equiv  { \sum\nolimits_{\alpha=1}^{\infty}  \delta (\mathbf{r}^\alpha - \mathbf{r}) f(\mathbf{r}^\alpha)}/
{\sum\nolimits_{\alpha=1}^{\infty}  \delta (\mathbf{r}^\alpha - \mathbf{r})}$ and 
$\hat{\mathcal{D}}_\mathbf{R} [g(\mathbf{R}^\alpha)] \equiv  { \sum\nolimits_{\alpha=1}^{\infty}  \delta (\mathbf{R}^\alpha - \mathbf{R}) g(\mathbf{R}^\alpha) }/
{\sum\nolimits_{\alpha=1}^{\infty}  \delta (\mathbf{R}^\alpha - \mathbf{R})}$, connecting respectively the (parametrized) electronic and nuclear subspaces with the full configuration space.  
Introducing the definitions (\ref{psin}) and (\ref{psie}) respectively into Eqs. (\ref{reconstruction_n}) and (\ref{reconstruction_e}), the full wave function 
$\Psi(\mathbf{r}, \mathbf{R} ,t)$ is immediately recovered provided that $\{ \mathbf{r}^\alpha,\mathbf{R}^\alpha \}$ exhaust the support of $|\Psi(\mathbf{r},\mathbf{R},t)|^2$. Notice that 
the second condition in (\ref{reconstruction_n}) and (\ref{reconstruction_e}) is required in order to avoid singularities due to the formation of nodes.

Part (b) Equations (\ref{cond_2}) and (\ref{cond_1}) can be derived by evaluating the Schr\"odinger equation (\ref{Scho}) at the configuration of the electronic and nuclear trajectories respectively, 
$\mathbf{r}^\alpha$ and $\mathbf{R}^\alpha$, and using the chain rule to write the time derivatives as
$d_t \psi_{n}  =  \partial_t \psi_{n}  +  \nabla_{\mathbf{r}}  \Psi  |_{\mathbf{r}^\alpha} \cdot \dot{\mathbf{r}}^\alpha$ and 
$d_t \psi_{e}  =  \partial_t \psi_{e}  +  \nabla_{\mathbf{R}}  \Psi  |_{\mathbf{R}^\alpha} \cdot \dot{\mathbf{R}}^\alpha$.

As written in (\ref{psin}) and (\ref{psie}), $\psi_{n}(\mathbf{R},t;\mathbf{r}^\alpha)$ and $\psi_{e}(\mathbf{r},t;\mathbf{R}^\alpha)$ represent $3N_n$- and $3N_e$-dimensional \textit{slices} 
of the full molecular wave function taken along the nuclear and electronic coordinates respectively. 
Each conditional wave function constitutes in this regard an open quantum system. Their evolution is non-unitary due to the last two terms 
in (\ref{cond_2}) and (\ref{cond_1}), in general complex functionals of the full wave function. 
The non-unitarity of Eqs. (\ref{cond_2}) and (\ref{cond_1}) is the result of separating a certain number of degrees of freedom without tracing over the rest.
From this point of view, the propagation of the nuclear equations of motion (\ref{cond_2}) does not require the calculation of BOPESs or NACs.
This makes the method particularly advantageous when studying processes that involve many BOPESs or external electromagnetic fields, as in laser-induced dynamics or 
scattering from metallic surfaces. 

Let us emphasize that the decomposition of the molecular wave function in (\ref{psin}) and (\ref{psie}) is only one case among many other possible conditional decompositions. 
The above theorem provides a general prescription to decompose the electron-nuclear wave function into a complete set of conditional wave functions. 
Particularly appealing is also the separation of the full wave function into single-particle conditional wave functions, i.e.  
$\psi^\nu_{n}(\mathbf{R}_\nu,t;\mathbf{R}^\alpha_1,..,\mathbf{R}^\alpha_{\nu-1},\mathbf{R}^\alpha_{\nu+1},..,\mathbf{r}^\alpha)$ and
$\psi^\xi_{e}(\mathbf{r}_\xi,t;\mathbf{r}^\alpha_1,..,\mathbf{r}^\alpha_{\xi-1},\mathbf{r}^\alpha_{\xi+1},..,\mathbf{R}^\alpha)$. 
Since the initial conditions of a trajectory-based simulation can be generated with importance sampling techniques, conditional decompositions allow to circumvent the problem of storing and propagating 
a many-particle wave function whose size scales exponentially with the number of particles. 

In the above theorem, it remains to specify the trajectories $\{\mathbf{r}^\alpha,\mathbf{R}^\alpha\}$. As already mentioned, the only requirement to be fulfilled by these trajectories is that 
they must explore the support of the quantum probability density $|\Psi(\mathbf{r},\mathbf{R},t)|^2$.
Notice that for the simplest case where $\dot{\mathbf{r}}^\alpha = \dot{\mathbf{R}}^\alpha = 0$, Eqs. (\ref{cond_2}) and (\ref{cond_1}) both reduce to the TDSE (\ref{Scho}). 
Alternatively, other choices of $\{\mathbf{r}^\alpha,\mathbf{R}^\alpha\}$ can be used to circumvent the use of computationally demanding fixed-grid methods.
Here we choose $\{\mathbf{r}^\alpha,\mathbf{R}^\alpha\}$ to be Bohmian trajectories because they do sample the quantum probability density \cite{BookOriols,*Holland} and 
because they provide in addition an intuitive picture of quantum dynamics \cite{STIRAP,*harmonic_gen1,*harmonic_gen2,*attosecond}.
Specifically, a proper sampling of the initial electron-nuclear wave function guarantees that $|\Psi(\mathbf{r},\mathbf{R},t)|^2 $ is exactly reproduced at any time by quantum trajectories 
$\left\{\mathbf{r}^{\alpha},\mathbf{R}^{\alpha} \right\}$ defined as
\begin{equation}\label{Traj_e}
  \mathbf{r}_\xi^\alpha = \mathbf{r}_\xi^\alpha + \int\nolimits_{t_0}^t \mathbf{v}_\xi^e(\mathbf{r}^\alpha(t'),\mathbf{R}^\alpha(t'), t')dt',
\end{equation}
\begin{equation}\label{Traj_n}
  \mathbf{R}_\nu^\alpha = \mathbf{R}_\nu^\alpha + \int\nolimits_{t_0}^t \mathbf{v}_\nu^n(\mathbf{r}^\alpha(t'),\mathbf{R}^\alpha(t'), t')dt',
\end{equation}
where electronic and nuclear velocity fields are defined as $\mathbf{v}_\xi^e (\mathbf{r},\mathbf{R},t) = (\mathbf{\nabla}_\xi S - \textbf{A}^{\mathcal{EM}})/{m}$ and 
$\mathbf{v}_\nu^n (\mathbf{r},\mathbf{R},t) = (\mathbf{\nabla}_\nu S - \textbf{A}^{\mathcal{EM}})/{M_\nu}$, and $S(\mathbf{r},\mathbf{R},t)$ is the phase of the full wave function 
$\Psi=|\Psi|e^{iS}$ \cite{BookOriols,*Holland}. Note that the choice of Bohmian trajectories is not mandatory. Alternatively, trajectory-based Monte-Carlo or importance-sampling techniques can be used 
provided that they sample the quantum-probability density. 

While not required in principle, in practice it is useful to propagate both the nuclear and electronic conditional wave functions, (\ref{cond_2}) and (\ref{cond_1}), to compute the quantum 
trajectories via conditional velocity fields defined as 
$\mathbf{v}_\xi^e (\mathbf{r}^\alpha,t;\mathbf{R}^\alpha) = (\nabla_\xi S_e(\mathbf{r},t;\mathbf{R}^\alpha) - \textbf{A}^{\mathcal{EM}}(\mathbf{r}_\xi))/{m}|_{\mathbf{r}^\alpha}$ and 
$\mathbf{v}_\nu^n (\mathbf{R}^\alpha,t;\mathbf{r}^\alpha) = (\nabla_\nu S_n(\mathbf{R},t;\mathbf{r}^\alpha) - \textbf{A}^{\mathcal{EM}}(\mathbf{R}_\nu))/{M_\nu}|_{\mathbf{R}^\alpha}$,
where $S_e(\mathbf{r},t;\mathbf{R}^\alpha)$ and $S_n(\mathbf{R},t;\mathbf{r}^\alpha)$ are respectively the phases of the electronic and nuclear conditional wave functions $\psi_n=|\psi_n|e^{iS_n}$ and 
$\psi_e=|\psi_e|e^{iS_e}$. 
In this way the reconstruction of the full wave function is avoided at the expense of solving twice the number of equations of motion \cite{PRLOriols,PRBGuille}.
Remarkably, the resulting propagation scheme, namely Eqs. (\ref{cond_2}) and (\ref{cond_1}) together with the trajectories in (\ref{Traj_e}) and (\ref{Traj_n}), 
does not require the computation of the quantum potential, in this manner overcoming a bottleneck in quantum trajectory-based approaches
\cite{Tavernelli3,*Wyatt3,*Meier1,*Prezhdo}.

In the remaining part of the letter, we explore a first approximation to this general method to solve the vibronic problem.
Let us first consider the external vector potential $\mathbf{A}^{\mathcal{EM}}$ to be zero.
In addition, we assume a zero order expansion of the complex functionals in (\ref{cond_2}) and (\ref{cond_1}) around the nuclear and electronic variables respectively, i.e.
${\hat T_e \Psi} |_{\mathbf{r}^\alpha} + i  {\mathbf{\nabla_{r}} \Psi} |_{\mathbf{r}^\alpha} \cdot \dot{\mathbf{r}}^\alpha = f_n(\mathbf{r}^\alpha,t)$ and
${\hat T_n \Psi} |_{\mathbf{R}^\alpha} + i  {\mathbf{\nabla_{R}} \Psi} |_{\mathbf{R}^\alpha} \cdot \dot{\mathbf{R}}^\alpha = f_e(\mathbf{R}^\alpha,t)$.
Notice that this approximation corresponds to the Hermitian limit of Eqs. (\ref{cond_2}) and (\ref{cond_1}) and thus that the time evolution of 
$\psi_{n}(\mathbf{R},t;\mathbf{r}^\alpha)$ and $\psi_{e}(\mathbf{r},t;\mathbf{R}^\alpha)$ becomes unitary.
The approximated functionals entail now a pure time-dependent phase that can be omitted because the velocity fields $v^n_\nu(\mathbf{r},\mathbf{R},t)$ and $v^e_\xi(\mathbf{r},\mathbf{R},t)$ 
are invariant under such a global phase transformation. We call the resulting propagation scheme, i.e. Eqs. (\ref{Traj_e}) and (\ref{Traj_n}) together with 
the Hermitian limit of Eqs. (\ref{cond_2}) and (\ref{cond_1}), the \textit{Hermitian conditional approach}.

To assess this approximated scheme, it is useful to restore the concept of potential-energy surface. 
This can be done by connecting this general method to the exact factorization of the molecular wave function \cite{Ali}. 
By rewriting the nuclear conditional wave function as a direct product of electronic and nuclear probability amplitudes, i.e. 
\begin{equation}\label{psin_ali}
\psi_{n}(\mathbf{R},t;\mathbf{r}^\alpha) 
= \Phi_{\mathbf{R}}(\mathbf{r}^\alpha,t) \chi(\mathbf{R},t),
\end{equation}
equations of motion for both terms in (\ref{psin_ali}) can be derived. Of particular interest is here $\chi(\mathbf{R},t)$ 
because it allows to isolate the role played by each term in (\ref{cond_2}) on the dynamics of the nuclear probability density in terms of a time-dependent potential-energy surface (TDPES) \cite{suppl1}.
In particular, the Hermitian limit of (\ref{cond_2}) leads to the following equation of motion for $\chi(\mathbf{R},t)$ \cite{suppl1}
\begin{eqnarray}\label{cond_nuc_approx}
i\partial_t \chi(\mathbf{R},t)   =   \Bigg\{  \sum\limits_{\nu=1}^{N_n}\frac{1}{2M_\nu}\big(  -i\nabla_\nu  +  \mathbf{\mathcal{A}}_\nu(\mathbf{R},t)  \big)^2  
\nonumber\\
+  W^n_{ext}(\mathbf{R},t)  +   \tilde\epsilon(\mathbf{R},t) \Bigg\}\chi(\mathbf{R},t),
\end{eqnarray}
where $\mathbf{\mathcal{A}}_\nu(\mathbf{R},t)$ is the $\nu-$component of the time-dependent Berry phase \cite{Ali}, and the approximated TDPES, $\tilde\epsilon(\mathbf{R},t)$, are defined as 
\begin{eqnarray}\label{TDPES_approx}
\tilde\epsilon(\mathbf{R},t) =  \epsilon(\mathbf{R},t) \nonumber\\
    - \int \hat{\mathcal{D}}_\mathbf{r} \left[\Phi^*_{\mathbf{R}}(\mathbf{r}^\alpha,t) 
       \left(( \hat T_e  + \dot{\mathbf{r}}^\alpha \cdot \nabla_{\mathbf{r}} ) \Phi_{\mathbf{R}}(\mathbf{r},t)\right)\Big{|}_{\mathbf{r}^\alpha} \right] d\mathbf{r}, \qquad
\end{eqnarray}
where $\Phi_{\mathbf{R}}(\mathbf{r}^\alpha,t) = \psi_{n}(\mathbf{R},t;\mathbf{r}^\alpha)/\chi(\mathbf{R},t)$.
Equation (\ref{cond_nuc_approx}) establishes a direct correspondence between the complex functionals in (\ref{cond_2}) and the last term on the r.h.s of (\ref{TDPES_approx}).
Note that computing the nuclear probability density from approximated nuclear conditional wave functions 
as $|\chi(\mathbf{R},t)|^2 = \int d\mathbf{r} \hat{\mathcal{D}}_\mathbf{r} \left[|\psi_{n}(\mathbf{R},t;\mathbf{r}^\alpha)\right|^2]$
is equivalent to propagate the nuclear probability density according to Eqs. (\ref{cond_nuc_approx}) and (\ref{TDPES_approx}) \cite{suppl1}.
Neglecting the averaged electronic kinetic energy in (\ref{TDPES_approx}) could seem a crude approximation in the 
Born-Oppenheimer limit \cite{Ali}, however in the following example we show that this is not the case when non-adiabatic effects are important.

\begin{figure}
\includegraphics[width=\columnwidth,height=6.75cm]{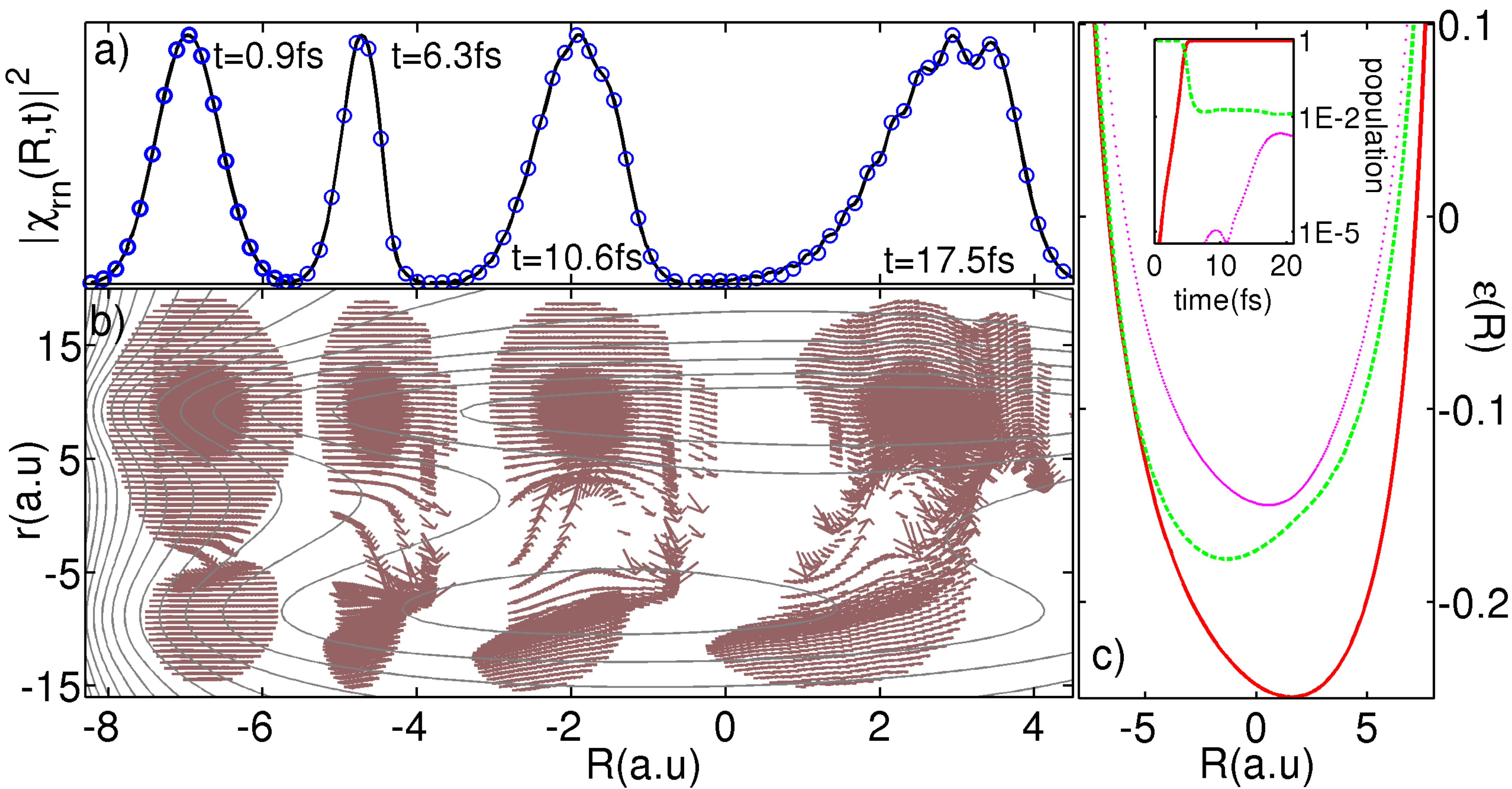}
\caption{(a) Exact (black solid-line) and approximated (blue circles) nuclear probability densities renormalized as $|\chi_{rn}(\mathbf{R},t)|^2 = |\chi(\mathbf{R},t)|^2/\max{|\chi(\mathbf{R},t)|^2}$  
at four different times. 
(b) Arrows refer to the (two-dimensional) velocity field computed from the approximated conditional wave \mbox{functions}. 
The gray contour lines represent the corresponding electron-nuclear two-dimensional potential energy surface. 
(c) First (red), second (green) and third (magenta) BOPESs involved in the non-adiabatic process. In the inset: adiabatic populations as a function of time computed from the exact solution.}
\label{fig}
\end{figure}

In what follows, we address two distinctive aspects of the correlated electron-nuclear motion, namely tunneling and interferences. 
A detailed discussion of the performance of the Hermitian conditional scheme to describe the splitting of the nuclear probability density can be found in \cite{suppl2}.
A numerically exactly solvable problem that exhibits the characteristic features associated with non-adiabatic processes is the model of Shin and Metiu \cite{Metiu}, which consists of three ions 
and a single electron. Two ions are fixed at a distance $L = 19.0a_0$, and the third ion and the electron are free to move in one dimension along the line joining the fixed ions. 
The Hamiltonian for this system reads
\begin{eqnarray}\label{metiu_ham}
&\hat H(r,R)& = -\frac{1}{2}\frac{\partial^2}{\partial r^2} - \frac{1}{2M}\frac{\partial^2}{\partial R^2} + \frac{1}{|\frac{L}{2} - R|} + \frac{1}{|\frac{L}{2} + R|} \nonumber\\
&-& \frac{erf\big(\frac{|R - r|}{R_f}\big)}{|R - r|}  - \frac{erf\big(\frac{|r - \frac{L}{2}|}{R_r}\big)}{|r - \frac{L}{2}|}  - \frac{erf\big(\frac{|r + \frac{L}{2}|}{R_l}\big)}{|r + \frac{L}{2}|},
\end{eqnarray}
where the symbols $\mathbf{r}$ and $\mathbf{R}$ are replaced by $r$ and $R$, and the coordinates of the electron and the movable nucleus are measured from the center of the two fixed ions. 
For the remaining parameters we choose $M = 1836$a.u. and $R_f = 7a_0$, $R_l = 4.4a_0$, and $R_r = 3.1a_0$ such that the first BOPES, $\epsilon^{(1)}_{BO}$, is strongly coupled to the second BOPES, $\epsilon^{(2)}_{BO}$, 
within an extended region defined by $R < -4a_0$. In addition, there is a moderate coupling between the second BOPES, $\epsilon^{(2)}_{BO}$, and the third BOPES, $\epsilon^{(3)}_{BO}$ for $R > 2a_0$ 
(see Fig.\ref{fig}.c). The coupling to the rest of the BOPESs is negligible.  
We suppose the system to be initially excited to $\epsilon^{(2)}_{BO}$ and the initial nuclear wave function to be a Gaussian wavepacket with  
$\sigma = 1/\sqrt{2.85}$, centered at $R = -7.0a_0$, i.e. the initial full wave function is $\Psi(r,R,t_0) = A e^{-(R+7)^2/\sigma^2} \Phi^{(2)}_R(r)$ with $A$ being a normalization constant. 
Starting with $\Psi(r,R,t_0)$, we first sample its probability density with trajectories and then propagate Eqs.~(\ref{Traj_e}) and (\ref{Traj_n}) together with the Hermitian limit of Eqs.~(\ref{cond_2}) and (\ref{cond_1}). 
In Fig.\ref{fig}.a we show snapshots at different times of the nuclear probability density for the exact calculation (in black solid line) and  
for the approximated solution (in blue circles) computed as 
$|\chi(\mathbf{R},t)|^2 = \int d\mathbf{r} \hat{\mathcal{D}}_\mathbf{r} \left[|\psi_{n}(\mathbf{R},t;\mathbf{r}^\alpha)\right|^2]$. 
Showing an excellent agreement, this propagation scheme is demonstrated to capture not only 
the conspicuous electronic transition between $\epsilon^{(2)}_{BO}$ and $\epsilon^{(1)}_{BO}$, but also the interferences originating at later times from 
contributions of higher adiabatic populations (see the rise of the population of $\epsilon^{(3)}_{BO}$ in the inset on Fig.\ref{fig}.c).

The BOPESs constitute a formidable interpretative tool to understand the electron-nuclear coupled dynamics, however they provide here a biased picture of the dynamics guiding the 
transit from the initial state at $t=0$fs to the final state at $t=17.5$fs. 
Alternatively, here we gain insight into this dynamics by analyzing the quantum velocity fields $\mathbf{v}^n(r,R,t)$ and $\mathbf{v}^e(r,R,t)$ computed respectively from the approximated conditional 
wave functions $\psi_{n}(\mathbf{R},t;\mathbf{r}^\alpha)$ and $\psi_{e}(\mathbf{r},t;\mathbf{R}^\alpha)$.
Snapshots of these velocity fields in terms of arrow maps are displayed in Fig.\ref{fig}.b together with contour lines representing the two-dimensional potential-energy surface. 
The first thing to notice is the fact that while in the picture of the BOPESs initial and final states are connected via tunneling (along the nuclear coordinates),
no tunneling is indeed taking place along this direction in the configuration space. 
At $t = 0.9$fs the trajectories at the rear of the wavepacket (with respect to the nuclear coordinates) carry a large momentum that forces 
the molecular wave function to squeeze at later times (e.g. at $t = 6.3$fs). 
This contraction in the nuclear coordinates is accompanied by a stretching of the wave function in the electronic coordinates that leads to a dripping of probability density out of the main ``reaction path'' via tunneling. 
During the tunneling process, quantum trajectories undergo a very fast motion in the electronic direction (notice the different sizes of the arrows in Fig.\ref{fig}.b). 
This induces a (tunneling) back and forth flow of probability density from one valley to the other 
(see the snapshots at times $t=10.6$fs and $t=17.5$fs). 
As a direct consequence, an interference pattern originates in close analogy with the quantum ``Bobsled effect'' described by McCullough and Wyatt \cite{bobsled_whirlpool1,*bobsled_whirlpool2}. 
A remarkably vortical behavior, reminiscent of the quantum ``whirlpool effect'' \cite{bobsled_whirlpool1,*bobsled_whirlpool2}, can be also observed when quasi-nodes in the
full wave function develop at $t=10.6$fs and $t=17.5$fs.  
This example demonstrates that the conditional formalism, even in the Hermitian limit, provides a powerful tool to describe complex features ubiquitous in non-adiabatic processes  
such as tunneling or interferences. Further, it evidences the interpretative value of the conditional formalism to grasp the ``microscopic'' behavior 
of quantum dynamics in terms of local velocities. 

To summarize, we present an exact trajectory-based decomposition of the Schr\"odinger equation in terms of conditional nuclear and electronic wave functions (\ref{psin}) and (\ref{psie}). 
Their evolution according to equations (\ref{cond_2}) and (\ref{cond_1}) is non-unitary and lends itself as a rigorous procedure to tackle open quantum systems by means of trajectory-based statistical 
techniques. 
In particular, the propagation of equation (\ref{cond_2}) does not entail integrals over the electronic degrees of freedom and 
hence we expect it to be of particular interest in scenarios where several BOPESs and external electromagnetic fields are involved. 
For an exactly solvable model system, even a zero order approximation is able to accurately reproduce complex non-adiabatic dynamics with quantum nuclear effects.   
The use of Bohmian trajectories adds interpretative value to the method and provides a numerically stable algorithm to avoid the calculation of the unstable quantum potential. 
Nonetheless, other kind of trajectories-based statistical techniques could be used as well. 
In this respect, the use of time-dependent density functional theory to sample the electronic subspace in combination with Eq. (\ref{cond_2}) is currently under study.

We gratefully acknowledge John C. Tully and Xavier Oriols for useful conversations. GA acknowledges the Beatriu de Pin\'os program for financial support through the project 2010BP-A00069.
IF thanks the Alexander von Humboldt Foundation for financial support. 
AR acknowledges support by the European Research Council Advanced Grant DYNamo (ERC-2010-AdG-267374), Spanish Grant (FIS2010- 21282-C02-01), Grupos Consolidados UPV/EHU del Gobierno Vasco (IT-578-13), 
Ikerbasque and European Commission project CRONOS (Grant number 280879- 2).

\bibliography{NAMD_CW_BIB}

\begin{thebibliography}{32}%
\makeatletter
\providecommand \@ifxundefined [1]{%
 \@ifx{#1\undefined}
}%
\providecommand \@ifnum [1]{%
 \ifnum #1\expandafter \@firstoftwo
 \else \expandafter \@secondoftwo
 \fi
}%
\providecommand \@ifx [1]{%
 \ifx #1\expandafter \@firstoftwo
 \else \expandafter \@secondoftwo
 \fi
}%
\providecommand \natexlab [1]{#1}%
\providecommand \enquote  [1]{``#1''}%
\providecommand \bibnamefont  [1]{#1}%
\providecommand \bibfnamefont [1]{#1}%
\providecommand \citenamefont [1]{#1}%
\providecommand \href@noop [0]{\@secondoftwo}%
\providecommand \href [0]{\begingroup \@sanitize@url \@href}%
\providecommand \@href[1]{\@@startlink{#1}\@@href}%
\providecommand \@@href[1]{\endgroup#1\@@endlink}%
\providecommand \@sanitize@url [0]{\catcode `\\12\catcode `\$12\catcode
  `\&12\catcode `\#12\catcode `\^12\catcode `\_12\catcode `\%12\relax}%
\providecommand \@@startlink[1]{}%
\providecommand \@@endlink[0]{}%
\providecommand \url  [0]{\begingroup\@sanitize@url \@url }%
\providecommand \@url [1]{\endgroup\@href {#1}{\urlprefix }}%
\providecommand \urlprefix  [0]{URL }%
\providecommand \Eprint [0]{\href }%
\providecommand \doibase [0]{http://dx.doi.org/}%
\providecommand \selectlanguage [0]{\@gobble}%
\providecommand \bibinfo  [0]{\@secondoftwo}%
\providecommand \bibfield  [0]{\@secondoftwo}%
\providecommand \translation [1]{[#1]}%
\providecommand \BibitemOpen [0]{}%
\providecommand \bibitemStop [0]{}%
\providecommand \bibitemNoStop [0]{.\EOS\space}%
\providecommand \EOS [0]{\spacefactor3000\relax}%
\providecommand \BibitemShut  [1]{\csname bibitem#1\endcsname}%
\let\auto@bib@innerbib\@empty
\bibitem [{\citenamefont {Tully}(2012)}]{TullyPerspective}%
  \BibitemOpen
  \bibfield  {author} {\bibinfo {author} {\bibfnamefont {J.~C.}\ \bibnamefont
  {Tully}},\ }\href
  {http://scitation.aip.org/content/aip/journal/jcp/137/22/10.1063/1.4757762}
  {\bibfield  {journal} {\bibinfo  {journal} {J. Chem. Phys.}\ }\textbf
  {\bibinfo {volume} {137}},\ \bibinfo {pages} {22A301} (\bibinfo {year}
  {2012})}\BibitemShut {NoStop}%
\bibitem [{\citenamefont {Domcke}\ and\ \citenamefont {Stock}(2007)}]{domcke}%
  \BibitemOpen
  \bibfield  {author} {\bibinfo {author} {\bibfnamefont {W.}~\bibnamefont
  {Domcke}}\ and\ \bibinfo {author} {\bibfnamefont {G.}~\bibnamefont {Stock}},\
  }\enquote {\bibinfo {title} {Theory of ultrafast nonadiabatic excited-state
  processes and their spectroscopic detection in real time},}\ in\ \href
  {http://dx.doi.org/10.1002/9780470141595.ch1} {\emph {\bibinfo {booktitle}
  {Adv. Chem. Phys.}}}\ (\bibinfo  {publisher} {John Wiley \& Sons, Inc.},\
  \bibinfo {year} {2007})\ pp.\ \bibinfo {pages} {1--169}\BibitemShut {NoStop}%
\bibitem [{\citenamefont {Beck}\ \emph {et~al.}(2000)\citenamefont {Beck},
  \citenamefont {Jäckle}, \citenamefont {Worth},\ and\ \citenamefont
  {Meyer}}]{MCTDH}%
  \BibitemOpen
  \bibfield  {author} {\bibinfo {author} {\bibfnamefont {M.}~\bibnamefont
  {Beck}}, \bibinfo {author} {\bibfnamefont {A.}~\bibnamefont {Jäckle}},
  \bibinfo {author} {\bibfnamefont {G.}~\bibnamefont {Worth}}, \ and\ \bibinfo
  {author} {\bibfnamefont {H.-D.}\ \bibnamefont {Meyer}},\ }\href
  {http://www.sciencedirect.com/science/article/pii/S0370157399000472}
  {\bibfield  {journal} {\bibinfo  {journal} {Phys. Rep.}\ }\textbf {\bibinfo
  {volume} {324}},\ \bibinfo {pages} {1 } (\bibinfo {year} {2000})}\BibitemShut
  {NoStop}%
\bibitem [{\citenamefont {Worth}\ and\ \citenamefont {Robb}(2003)}]{robb}%
  \BibitemOpen
  \bibfield  {author} {\bibinfo {author} {\bibfnamefont {G.~A.}\ \bibnamefont
  {Worth}}\ and\ \bibinfo {author} {\bibfnamefont {M.~A.}\ \bibnamefont
  {Robb}},\ }\enquote {\bibinfo {title} {Applying direct molecular dynamics to
  non-adiabatic systems},}\ in\ \href
  {http://dx.doi.org/10.1002/0471433462.ch7} {\emph {\bibinfo {booktitle} {The
  Role of Degenerate States in Chemistry}}}\ (\bibinfo  {publisher} {John Wiley
  \& Sons, Inc.},\ \bibinfo {year} {2003})\ pp.\ \bibinfo {pages}
  {355--431}\BibitemShut {NoStop}%
\bibitem [{\citenamefont {Truhlar}(1995)}]{truhlar}%
  \BibitemOpen
  \bibfield  {author} {\bibinfo {author} {\bibfnamefont {D.}~\bibnamefont
  {Truhlar}},\ }in\ \href {http://dx.doi.org/10.1007/978-94-015-8539-2_10}
  {\emph {\bibinfo {booktitle} {The Reaction Path in Chemistry: Current
  Approaches and Perspectives}}},\ \bibinfo {series} {Understanding Chemical
  Reactivity}, Vol.~\bibinfo {volume} {16},\ \bibinfo {editor} {edited by\
  \bibinfo {editor} {\bibfnamefont {D.}~\bibnamefont {Heidrich}}}\ (\bibinfo
  {publisher} {Springer Netherlands},\ \bibinfo {year} {1995})\ pp.\ \bibinfo
  {pages} {229--255}\BibitemShut {NoStop}%
\bibitem [{\citenamefont {C.~Tully}(1998)}]{tully1998mixed}%
  \BibitemOpen
  \bibfield  {author} {\bibinfo {author} {\bibfnamefont {J.}~\bibnamefont
  {C.~Tully}},\ }\href {http://dx.doi.org/10.1039/A801824C} {\bibfield
  {journal} {\bibinfo  {journal} {Farad. Discuss.}\ }\textbf {\bibinfo {volume}
  {110}},\ \bibinfo {pages} {407} (\bibinfo {year} {1998})}\BibitemShut
  {NoStop}%
\bibitem [{\citenamefont {Sawada}\ and\ \citenamefont
  {Metiu}(1986)}]{SawadaMetiu}%
  \BibitemOpen
  \bibfield  {author} {\bibinfo {author} {\bibfnamefont {S.}~\bibnamefont
  {Sawada}}\ and\ \bibinfo {author} {\bibfnamefont {H.}~\bibnamefont {Metiu}},\
  }\href
  {http://scitation.aip.org/content/aip/journal/jcp/84/1/10.1063/1.450175}
  {\bibfield  {journal} {\bibinfo  {journal} {J. Chem. Phys.}\ }\textbf
  {\bibinfo {volume} {84}},\ \bibinfo {pages} {227} (\bibinfo {year}
  {1986})}\BibitemShut {NoStop}%
\bibitem [{\citenamefont {Burghardt}\ \emph {et~al.}(1999)\citenamefont
  {Burghardt}, \citenamefont {Meyer},\ and\ \citenamefont
  {Cederbaum}}]{BurghardtCederbaum}%
  \BibitemOpen
  \bibfield  {author} {\bibinfo {author} {\bibfnamefont {I.}~\bibnamefont
  {Burghardt}}, \bibinfo {author} {\bibfnamefont {H.-D.}\ \bibnamefont
  {Meyer}}, \ and\ \bibinfo {author} {\bibfnamefont {L.~S.}\ \bibnamefont
  {Cederbaum}},\ }\href
  {http://scitation.aip.org/content/aip/journal/jcp/111/7/10.1063/1.479574}
  {\bibfield  {journal} {\bibinfo  {journal} {J. Chem. Phys.}\ }\textbf
  {\bibinfo {volume} {111}},\ \bibinfo {pages} {2927} (\bibinfo {year}
  {1999})}\BibitemShut {NoStop}%
\bibitem [{\citenamefont {Martinez}\ \emph {et~al.}(1996)\citenamefont
  {Martinez}, \citenamefont {Ben-Nun},\ and\ \citenamefont
  {Levine}}]{Martinez}%
  \BibitemOpen
  \bibfield  {author} {\bibinfo {author} {\bibfnamefont {T.~J.}\ \bibnamefont
  {Martinez}}, \bibinfo {author} {\bibfnamefont {M.}~\bibnamefont {Ben-Nun}}, \
  and\ \bibinfo {author} {\bibfnamefont {R.~D.}\ \bibnamefont {Levine}},\
  }\href {http://pubs.acs.org/doi/abs/10.1021/jp953105a} {\bibfield  {journal}
  {\bibinfo  {journal} {J. Phys. Chem.}\ }\textbf {\bibinfo {volume} {100}},\
  \bibinfo {pages} {7884} (\bibinfo {year} {1996})}\BibitemShut {NoStop}%
\bibitem [{\citenamefont {Sanchez-Galvez}\ \emph {et~al.}(2000)\citenamefont
  {Sanchez-Galvez}, \citenamefont {Hunt}, \citenamefont {Robb}, \citenamefont
  {Olivucci}, \citenamefont {Vreven},\ and\ \citenamefont
  {Schlegel}}]{results1}%
  \BibitemOpen
  \bibfield  {author} {\bibinfo {author} {\bibfnamefont {A.}~\bibnamefont
  {Sanchez-Galvez}}, \bibinfo {author} {\bibfnamefont {P.}~\bibnamefont
  {Hunt}}, \bibinfo {author} {\bibfnamefont {M.~A.}\ \bibnamefont {Robb}},
  \bibinfo {author} {\bibfnamefont {M.}~\bibnamefont {Olivucci}}, \bibinfo
  {author} {\bibfnamefont {T.}~\bibnamefont {Vreven}}, \ and\ \bibinfo {author}
  {\bibfnamefont {H.~B.}\ \bibnamefont {Schlegel}},\ }\href
  {http://pubs.acs.org/doi/abs/10.1021/ja993985x} {\bibfield  {journal}
  {\bibinfo  {journal} {J. Am. Chem. Soc.}\ }\textbf {\bibinfo {volume}
  {122}},\ \bibinfo {pages} {2911} (\bibinfo {year} {2000})}\BibitemShut
  {NoStop}%
\bibitem [{\citenamefont {Jolibois}\ \emph {et~al.}(2000)\citenamefont
  {Jolibois}, \citenamefont {Bearpark}, \citenamefont {Klein}, \citenamefont
  {Olivucci},\ and\ \citenamefont {Robb}}]{results2}%
  \BibitemOpen
  \bibfield  {author} {\bibinfo {author} {\bibfnamefont {F.}~\bibnamefont
  {Jolibois}}, \bibinfo {author} {\bibfnamefont {M.~J.}\ \bibnamefont
  {Bearpark}}, \bibinfo {author} {\bibfnamefont {S.}~\bibnamefont {Klein}},
  \bibinfo {author} {\bibfnamefont {M.}~\bibnamefont {Olivucci}}, \ and\
  \bibinfo {author} {\bibfnamefont {M.~A.}\ \bibnamefont {Robb}},\ }\href
  {http://pubs.acs.org/doi/abs/10.1021/ja992717w} {\bibfield  {journal}
  {\bibinfo  {journal} {J. Am. Chem. Soc.}\ }\textbf {\bibinfo {volume}
  {122}},\ \bibinfo {pages} {5801} (\bibinfo {year} {2000})}\BibitemShut
  {NoStop}%
\bibitem [{\citenamefont {Ben-Nun}\ \emph {et~al.}(1998)\citenamefont
  {Ben-Nun}, \citenamefont {Molnar}, \citenamefont {Lu}, \citenamefont
  {C.~Phillips}, \citenamefont {J.~Martinez},\ and\ \citenamefont
  {Schulten}}]{results3}%
  \BibitemOpen
  \bibfield  {author} {\bibinfo {author} {\bibfnamefont {M.}~\bibnamefont
  {Ben-Nun}}, \bibinfo {author} {\bibfnamefont {F.}~\bibnamefont {Molnar}},
  \bibinfo {author} {\bibfnamefont {H.}~\bibnamefont {Lu}}, \bibinfo {author}
  {\bibfnamefont {J.}~\bibnamefont {C.~Phillips}}, \bibinfo {author}
  {\bibfnamefont {T.}~\bibnamefont {J.~Martinez}}, \ and\ \bibinfo {author}
  {\bibfnamefont {K.}~\bibnamefont {Schulten}},\ }\href
  {http://dx.doi.org/10.1039/A801310A} {\bibfield  {journal} {\bibinfo
  {journal} {Farad. Discuss.}\ }\textbf {\bibinfo {volume} {110}},\ \bibinfo
  {pages} {447} (\bibinfo {year} {1998})}\BibitemShut {NoStop}%
\bibitem [{\citenamefont {Oriols}(2007)}]{PRLOriols}%
  \BibitemOpen
  \bibfield  {author} {\bibinfo {author} {\bibfnamefont {X.}~\bibnamefont
  {Oriols}},\ }\href {http://link.aps.org/doi/10.1103/PhysRevLett.98.066803}
  {\bibfield  {journal} {\bibinfo  {journal} {Phys. Rev. Lett.}\ }\textbf
  {\bibinfo {volume} {98}},\ \bibinfo {pages} {066803} (\bibinfo {year}
  {2007})}\BibitemShut {NoStop}%
\bibitem [{\citenamefont {Albareda}\ \emph {et~al.}(2009)\citenamefont
  {Albareda}, \citenamefont {Su\~n\'e},\ and\ \citenamefont
  {Oriols}}]{PRBGuille}%
  \BibitemOpen
  \bibfield  {author} {\bibinfo {author} {\bibfnamefont {G.}~\bibnamefont
  {Albareda}}, \bibinfo {author} {\bibfnamefont {J.}~\bibnamefont {Su\~n\'e}},
  \ and\ \bibinfo {author} {\bibfnamefont {X.}~\bibnamefont {Oriols}},\ }\href
  {http://link.aps.org/doi/10.1103/PhysRevB.79.075315} {\bibfield  {journal}
  {\bibinfo  {journal} {Phys. Rev. B}\ }\textbf {\bibinfo {volume} {79}},\
  \bibinfo {pages} {075315} (\bibinfo {year} {2009})}\BibitemShut {NoStop}%
\bibitem [{\citenamefont {Albareda}\ \emph {et~al.}(2013)\citenamefont
  {Albareda}, \citenamefont {Marian}, \citenamefont {Benali}, \citenamefont
  {Yaro}, \citenamefont {Zanghì},\ and\ \citenamefont {Oriols}}]{JCE}%
  \BibitemOpen
  \bibfield  {author} {\bibinfo {author} {\bibfnamefont {G.}~\bibnamefont
  {Albareda}}, \bibinfo {author} {\bibfnamefont {D.}~\bibnamefont {Marian}},
  \bibinfo {author} {\bibfnamefont {A.}~\bibnamefont {Benali}}, \bibinfo
  {author} {\bibfnamefont {S.}~\bibnamefont {Yaro}}, \bibinfo {author}
  {\bibfnamefont {N.}~\bibnamefont {Zanghì}}, \ and\ \bibinfo {author}
  {\bibfnamefont {X.}~\bibnamefont {Oriols}},\ }\href {\doibase
  10.1007/s10825-013-0484-5} {\bibfield  {journal} {\bibinfo  {journal} {J.
  Comp. Electr.}\ }\textbf {\bibinfo {volume} {12}},\ \bibinfo {pages} {405}
  (\bibinfo {year} {2013})}\BibitemShut {NoStop}%
\bibitem [{\citenamefont {Albareda}\ \emph {et~al.}(2012)\citenamefont
  {Albareda}, \citenamefont {Traversa}, \citenamefont {Benali},\ and\
  \citenamefont {Oriols}}]{FNLGuille}%
  \BibitemOpen
  \bibfield  {author} {\bibinfo {author} {\bibfnamefont {G.}~\bibnamefont
  {Albareda}}, \bibinfo {author} {\bibfnamefont {F.~L.}\ \bibnamefont
  {Traversa}}, \bibinfo {author} {\bibfnamefont {A.}~\bibnamefont {Benali}}, \
  and\ \bibinfo {author} {\bibfnamefont {X.}~\bibnamefont {Oriols}},\ }\href
  {http://www.worldscientific.com/doi/abs/10.1142/S0219477512420084} {\bibfield
   {journal} {\bibinfo  {journal} {Fluct. Noise Lett.}\ }\textbf {\bibinfo
  {volume} {11}},\ \bibinfo {pages} {1242008} (\bibinfo {year}
  {2012})}\BibitemShut {NoStop}%
\bibitem [{\citenamefont {Oriols}\ and\ \citenamefont
  {Mompart}(2012)}]{BookOriols}%
  \BibitemOpen
  \bibfield  {author} {\bibinfo {author} {\bibfnamefont {X.}~\bibnamefont
  {Oriols}}\ and\ \bibinfo {author} {\bibfnamefont {J.}~\bibnamefont
  {Mompart}},\ }\href {http://books.google.de/books?id=mnqNx66amcIC} {\emph
  {\bibinfo {title} {Applied Bohmian Mechanics: From Nanoscale Systems to
  Cosmology}}}\ (\bibinfo  {publisher} {Pan Stanford},\ \bibinfo {year}
  {2012})\BibitemShut {NoStop}%
\bibitem [{\citenamefont {Holland}(1995)}]{Holland}%
  \BibitemOpen
  \bibfield  {author} {\bibinfo {author} {\bibfnamefont {P.}~\bibnamefont
  {Holland}},\ }\href {http://books.google.de/books?id=BsEfVBzToRMC} {\emph
  {\bibinfo {title} {The Quantum Theory of Motion: An Account of the de
  Broglie-Bohm Causal Interpretation of Quantum Mechanics}}}\ (\bibinfo
  {publisher} {Cambridge University Press},\ \bibinfo {year}
  {1995})\BibitemShut {NoStop}%
\bibitem [{\citenamefont {Benseny}\ \emph {et~al.}(2012)\citenamefont
  {Benseny}, \citenamefont {Bagud\`a}, \citenamefont {Oriols},\ and\
  \citenamefont {Mompart}}]{STIRAP}%
  \BibitemOpen
  \bibfield  {author} {\bibinfo {author} {\bibfnamefont {A.}~\bibnamefont
  {Benseny}}, \bibinfo {author} {\bibfnamefont {J.}~\bibnamefont {Bagud\`a}},
  \bibinfo {author} {\bibfnamefont {X.}~\bibnamefont {Oriols}}, \ and\ \bibinfo
  {author} {\bibfnamefont {J.}~\bibnamefont {Mompart}},\ }\href
  {http://link.aps.org/doi/10.1103/PhysRevA.85.053619} {\bibfield  {journal}
  {\bibinfo  {journal} {Phys. Rev. A}\ }\textbf {\bibinfo {volume} {85}},\
  \bibinfo {pages} {053619} (\bibinfo {year} {2012})}\BibitemShut {NoStop}%
\bibitem [{\citenamefont {Wu}\ \emph {et~al.}(2013)\citenamefont {Wu},
  \citenamefont {Augstein},\ and\ \citenamefont {Figueira~de
  Morisson~Faria}}]{harmonic_gen1}%
  \BibitemOpen
  \bibfield  {author} {\bibinfo {author} {\bibfnamefont {J.}~\bibnamefont
  {Wu}}, \bibinfo {author} {\bibfnamefont {B.~B.}\ \bibnamefont {Augstein}}, \
  and\ \bibinfo {author} {\bibfnamefont {C.}~\bibnamefont {Figueira~de
  Morisson~Faria}},\ }\href
  {http://link.aps.org/doi/10.1103/PhysRevA.88.023415} {\bibfield  {journal}
  {\bibinfo  {journal} {Phys. Rev. A}\ }\textbf {\bibinfo {volume} {88}},\
  \bibinfo {pages} {023415} (\bibinfo {year} {2013})}\BibitemShut {NoStop}%
\bibitem [{\citenamefont {Song}\ \emph {et~al.}(2013)\citenamefont {Song},
  \citenamefont {Li}, \citenamefont {Liu}, \citenamefont {Guo},\ and\
  \citenamefont {Yang}}]{harmonic_gen2}%
  \BibitemOpen
  \bibfield  {author} {\bibinfo {author} {\bibfnamefont {Y.}~\bibnamefont
  {Song}}, \bibinfo {author} {\bibfnamefont {S.-Y.}\ \bibnamefont {Li}},
  \bibinfo {author} {\bibfnamefont {X.-S.}\ \bibnamefont {Liu}}, \bibinfo
  {author} {\bibfnamefont {F.-M.}\ \bibnamefont {Guo}}, \ and\ \bibinfo
  {author} {\bibfnamefont {Y.-J.}\ \bibnamefont {Yang}},\ }\href
  {http://link.aps.org/doi/10.1103/PhysRevA.88.053419} {\bibfield  {journal}
  {\bibinfo  {journal} {Phys. Rev. A}\ }\textbf {\bibinfo {volume} {88}},\
  \bibinfo {pages} {053419} (\bibinfo {year} {2013})}\BibitemShut {NoStop}%
\bibitem [{\citenamefont {Takemoto}\ and\ \citenamefont
  {Becker}(2011)}]{attosecond}%
  \BibitemOpen
  \bibfield  {author} {\bibinfo {author} {\bibfnamefont {N.}~\bibnamefont
  {Takemoto}}\ and\ \bibinfo {author} {\bibfnamefont {A.}~\bibnamefont
  {Becker}},\ }\href
  {http://scitation.aip.org/content/aip/journal/jcp/134/7/10.1063/1.3553178}
  {\bibfield  {journal} {\bibinfo  {journal} {J. Chem. Phys.}\ }\textbf
  {\bibinfo {volume} {134}},\ \bibinfo {eid} {074309} (\bibinfo {year}
  {2011})}\BibitemShut {NoStop}%
\bibitem [{\citenamefont {Tavernelli}(2013)}]{Tavernelli3}%
  \BibitemOpen
  \bibfield  {author} {\bibinfo {author} {\bibfnamefont {I.}~\bibnamefont
  {Tavernelli}},\ }\href {\doibase 10.1103/PhysRevA.87.042501} {\bibfield
  {journal} {\bibinfo  {journal} {Phys. Rev. A}\ }\textbf {\bibinfo {volume}
  {87}},\ \bibinfo {pages} {042501} (\bibinfo {year} {2013})}\BibitemShut
  {NoStop}%
\bibitem [{\citenamefont {Lopreore}\ and\ \citenamefont
  {Wyatt}(2002)}]{Wyatt3}%
  \BibitemOpen
  \bibfield  {author} {\bibinfo {author} {\bibfnamefont {C.~L.}\ \bibnamefont
  {Lopreore}}\ and\ \bibinfo {author} {\bibfnamefont {R.~E.}\ \bibnamefont
  {Wyatt}},\ }\href
  {http://scitation.aip.org/content/aip/journal/jcp/116/4/10.1063/1.1427916}
  {\bibfield  {journal} {\bibinfo  {journal} {J. Chem. Phys.}\ }\textbf
  {\bibinfo {volume} {116}},\ \bibinfo {pages} {1228} (\bibinfo {year}
  {2002})}\BibitemShut {NoStop}%
\bibitem [{\citenamefont {Gindensperger}\ \emph {et~al.}(2000)\citenamefont
  {Gindensperger}, \citenamefont {Meier},\ and\ \citenamefont
  {Beswick}}]{Meier1}%
  \BibitemOpen
  \bibfield  {author} {\bibinfo {author} {\bibfnamefont {E.}~\bibnamefont
  {Gindensperger}}, \bibinfo {author} {\bibfnamefont {C.}~\bibnamefont
  {Meier}}, \ and\ \bibinfo {author} {\bibfnamefont {J.~A.}\ \bibnamefont
  {Beswick}},\ }\href
  {http://scitation.aip.org/content/aip/journal/jcp/113/21/10.1063/1.1328759",}
  {\bibfield  {journal} {\bibinfo  {journal} {J. Chem. Phys.}\ }\textbf
  {\bibinfo {volume} {113}},\ \bibinfo {pages} {9369} (\bibinfo {year}
  {2000})}\BibitemShut {NoStop}%
\bibitem [{\citenamefont {Prezhdo}\ and\ \citenamefont
  {Brooksby}(2001)}]{Prezhdo}%
  \BibitemOpen
  \bibfield  {author} {\bibinfo {author} {\bibfnamefont {O.~V.}\ \bibnamefont
  {Prezhdo}}\ and\ \bibinfo {author} {\bibfnamefont {C.}~\bibnamefont
  {Brooksby}},\ }\href {http://link.aps.org/doi/10.1103/PhysRevLett.86.3215}
  {\bibfield  {journal} {\bibinfo  {journal} {Phys. Rev. Lett.}\ }\textbf
  {\bibinfo {volume} {86}},\ \bibinfo {pages} {3215} (\bibinfo {year}
  {2001})}\BibitemShut {NoStop}%
\bibitem [{\citenamefont {Abedi}\ \emph {et~al.}(2010)\citenamefont {Abedi},
  \citenamefont {Maitra},\ and\ \citenamefont {Gross}}]{Ali}%
  \BibitemOpen
  \bibfield  {author} {\bibinfo {author} {\bibfnamefont {A.}~\bibnamefont
  {Abedi}}, \bibinfo {author} {\bibfnamefont {N.~T.}\ \bibnamefont {Maitra}}, \
  and\ \bibinfo {author} {\bibfnamefont {E.~K.~U.}\ \bibnamefont {Gross}},\
  }\href {http://link.aps.org/doi/10.1103/PhysRevLett.105.123002} {\bibfield
  {journal} {\bibinfo  {journal} {Phys. Rev. Lett.}\ }\textbf {\bibinfo
  {volume} {105}},\ \bibinfo {pages} {123002} (\bibinfo {year}
  {2010})}\BibitemShut {NoStop}%
\bibitem [{({\natexlab{a}})}]{suppl1}%
  \BibitemOpen
  \href@noop {} {} ({\natexlab{a}}),\ \bibinfo {note} {see Appendix A in the SM
  for a detailed derivation of the connection between the conditional formalism
  and the exact factorization of the molecular wave function}\BibitemShut
  {NoStop}%
\bibitem [{({\natexlab{b}})}]{suppl2}%
  \BibitemOpen
  \href@noop {} {} ({\natexlab{b}}),\ \bibinfo {note} {see Appendix B in the SM
  for a detailed discussion of the performance of the Hermitian conditional
  scheme to describe the splitting of the nuclear wavepacket}\BibitemShut
  {NoStop}%
\bibitem [{\citenamefont {Shin}\ and\ \citenamefont {Metiu}(1995)}]{Metiu}%
  \BibitemOpen
  \bibfield  {author} {\bibinfo {author} {\bibfnamefont {S.}~\bibnamefont
  {Shin}}\ and\ \bibinfo {author} {\bibfnamefont {H.}~\bibnamefont {Metiu}},\
  }\href
  {http://scitation.aip.org/content/aip/journal/jcp/102/23/10.1063/1.468795}
  {\bibfield  {journal} {\bibinfo  {journal} {J. Chem. Phys.}\ }\textbf
  {\bibinfo {volume} {102}},\ \bibinfo {pages} {9285} (\bibinfo {year}
  {1995})}\BibitemShut {NoStop}%
\bibitem [{\citenamefont {McCullough}\ and\ \citenamefont
  {Wyatt}(1969)}]{bobsled_whirlpool1}%
  \BibitemOpen
  \bibfield  {author} {\bibinfo {author} {\bibfnamefont {E.~A.}\ \bibnamefont
  {McCullough}}\ and\ \bibinfo {author} {\bibfnamefont {R.~E.}\ \bibnamefont
  {Wyatt}},\ }\href
  {http://scitation.aip.org/content/aip/journal/jcp/51/3/10.1063/1.1672133}
  {\bibfield  {journal} {\bibinfo  {journal} {J. Chem. Phys.}\ }\textbf
  {\bibinfo {volume} {51}},\ \bibinfo {pages} {1253} (\bibinfo {year}
  {1969})}\BibitemShut {NoStop}%
\bibitem [{\citenamefont {McCullough}\ and\ \citenamefont
  {Wyatt}(1971)}]{bobsled_whirlpool2}%
  \BibitemOpen
  \bibfield  {author} {\bibinfo {author} {\bibfnamefont {E.~A.}\ \bibnamefont
  {McCullough}}\ and\ \bibinfo {author} {\bibfnamefont {R.~E.}\ \bibnamefont
  {Wyatt}},\ }\href
  {http://scitation.aip.org/content/aip/journal/jcp/54/8/10.1063/1.1675384}
  {\bibfield  {journal} {\bibinfo  {journal} {J. Chem. Phys.}\ }\textbf
  {\bibinfo {volume} {54}},\ \bibinfo {pages} {3578} (\bibinfo {year}
  {1971})}\BibitemShut {NoStop}%
\end{thebibliography}%

\end{document}